\documentclass [a4paper,fleqn, 12pt]{article}
\usepackage{graphicx}

\usepackage {amsmath} \usepackage{amssymb} \usepackage{cite}

\newcommand{\cn}

\begin{document}


\title
{Nonlinear evolution equations for describing waves in bubbly liquids with viscosity and heat transfer consideration}

\author
{Nikolay A. Kudryashov, \and Dmitry I. Sinelshchikov}

\date{Department of Applied Mathematics, National Research Nuclear University
MEPHI, 31 Kashirskoe Shosse,
115409 Moscow, Russian Federation}




\maketitle

\begin{abstract}

Nonlinear evolution equations of the fourth order and its partial cases are derived for describing nonlinear pressure waves in a mixture liquid and gas bubbles. Influence of viscosity and heat transfer is taken into account. Exact solutions of nonlinear evolution equation of the fourth order are found by means of the simplest equation method. Properties of nonlinear waves in a liquid with gas bubbles are discussed.

\end{abstract}






\section{Introduction}

It is well known that a mixture of liquid and gas bubbles can be considered as an example of a nonlinear medium. Van Wijngarden \cite{Wijngaarden} was first who derived that the famous Korteweg -- de Vries equation \cite{Korteweg, Whitham, Drazin} can be used for describing nonlinear waves in liquid with gas bubbles. Taking into account the viscosity of liquid,  Nakoryakov and et al.\cite{Nakoryakov_1968} obtained the Burgers equation \cite{Bateman, Burgers} and the Korteweg - de Vries - Burgers equation to describe the pressure waves in bubbly liquids.
The dynamic propagation of acoustic waves in a half - space filled with a viscous,  bubbly liquid under van Wijngaarden linear theory was considered in the recent work \cite{Jordan2006}.
Oganyan in \cite{Oganyan_2005} investigated a quasi -- isothermal wave propagation in a gas -- liquid mixture, in which he take into consideration the thermodynamic behavior of the near -- isothermal  gas in the bubbles.

Many authors studied nonlinear processes in a liquid with gas bubbles using the numerical methods. For an example Nigmatullin and Khabeev \cite{Khabeev1974, Khabeev1977} considered the heat transfer between an gas bubble and a liquid and studied the structure of shock waves in a liquid with gas bubbles with consideration for the heat transfer.

The purpose of this work is to obtain nonlinear evolution equations for describing the pressure waves of in a liquid with gas bubbles taking into account the viscosity of liquid and the heat transfer on boundary a bubble and a liquid.

\section{System of equations for describing waves in a liquid with gas bubbles with consideration for heat transfer and viscosity}

Suppose that a mixture of a liquid and gas bubbles is homogeneous medium \cite{Nakoryakov_1972, Nigmatullin}. In this case for description of this mixture we use the averaged temperature, velocity, density and  pressure. We also assume that the gas bubbles has the same size and the amount of bubbles in the mass unit is constant. We take the processes of the heat transfer and viscosity on the boundary of bubble and liquid into account. We do not consider the processes of formation, destruction, and conglutination for the gas bubbles.

We have that the volume and the mass of gas in the unit of the mass mixture can be written as
\begin{equation*}
V=\frac{4}{3}\pi R^{3}N,\,\,\,\,\,X=V\rho_{g},
\end{equation*}
where $R=R(x,t)$ is bubble radius, $N$ is number of bubbles in mass unit, $\rho_{g}=\rho_{g}(x,t)$ is the gas density. Here and later we believe that the subscript $g$ corresponds to the gas phase and
subscript $l$ corresponds to the liquid phase.

We consider the long wavelength perturbations in a mixture of the liquid and the gas bubbles  assuming that characteristic length of waves of perturbation more than distance between bubbles.
We also assume, that distance between bubbles much more than the averaged radius of a bubble.

We describe dynamics of a bubble using the Rayleigh -- Lamb equation. We also take the equation of energy for a bubble and the state equation for the gas in a bubble into account. The system of equation for the description of the gas bubble takes the form \cite{Nakoryakov_1972, Nigmatullin, Rayleigh, Plesset, Wijng}
\begin{equation}
\rho_{l}\left(R R_{tt}+\frac{3}{2} R_{t}^{2}+\frac{4\nu}{3R}
R_{t}\right)=P_{g}-P, \label{eq: Relei}
\end{equation}

\begin{equation}
P_{g,\,t}+\frac{3nP_{g}}{R}R_{t}+\frac{3\chi_{g}\,Nu\,(n-1)}{2R^{2}}(T_{g}-T_{l})=0,
\label{eq: Energy}
\end{equation}

\begin{equation}
T_{g}=\frac{T_{0}P_{g}}{P_{g,\,0}}\left(\frac{R}{R_{0}}\right)^{3},
\label{eq: K-M}
\end{equation}
where  $P(x,t)$ is a pressure of a gas-liquid mixture, $P_{g}$ is a gas pressure in a bubble, $T_{g}$ and $T_{l}$ are temperatures of liquid and gas accordingly, $\chi_{g}$ is a coefficient of the gas thermal conduction, $Nu$ is the Nusselt number, $n$ is a polytropic exponent, $\nu$ is the viscosity of a liquid.

The expression for the density of a mixture can be presented in the form \cite{Nakoryakov_1968}
\begin{equation}
\frac{1}{\rho}=\frac{1-X}{\rho_{l}}+V\,\Rightarrow \rho=\frac{\rho_{l}}{1-X+V\rho_{l}}.
\label{eq: density}
\end{equation}

Considering the small deviation of the bubble radius in comparison with the averaged radius of bubble,  we have
\begin{equation}
  \begin{gathered}
 R(x,t) = R_{0} + \eta(x,t),\:\   \quad
  R_{0}=const,\:\   \quad ||\eta||<<R_{0},\quad
  R(x,0)=R_{0}.
  \end{gathered}
  \label{eq: rel1}
\end{equation}

Assume that the liquid temperature is constant and equal to the initial value
\begin{equation*}
T_{l}=T\mid_{{t=0}}=T_{0},\,\,\,\,\,T_{0}=const. \label{eq: T_eq}
\end{equation*}

At the initial moment, we also have
\begin{equation*}
  \begin{gathered}
   t=0: \quad P=P_{g}=P_{0},\,\,\,\,\,P_{0}=const,
    \quad V=V_{0}=\frac{4}{3}\pi R_{0}^{3}N.
   \end{gathered}
\end{equation*}

Substituting $P_g$ and $T_g$ from Eqs. (\ref{eq: Relei}) and (\ref{eq: K-M}) into Eq. (\ref{eq:
Energy}) and taking relation (\ref{eq: rel1}) into account we have the pressure dependence of a mixture on the radius perturbation in the form
\begin{equation}
  \begin{gathered}
P-P_{0}+\frac{\eta}{R_{0}}P+\frac{3n\varkappa}{R_{0}}P
\eta_{t}+\varkappa P_{t}+\varkappa \rho_{l}(R_{0}+\eta)
\eta_{ttt}+(3n+4) \varkappa \rho_{l} \eta_{t} \eta_{tt}+\\
\\+
\frac{\rho_{l}(3R_{0}^{2}+4\nu\varkappa)}{3R_{0}}\eta_{tt}+
\frac{\rho_{l}(6R_{0}^{2}-4\nu\varkappa)}{3R_{0}^{2}}\eta
\eta_{tt}+
\frac{\rho_{l}(8\nu\varkappa(3n-1)+9R_{0}^{2})}{6R_{0}^{2}}\eta_{t}^{2}+\\
\\+
\frac{4\nu\rho_{l}}{3R_{0}}\eta_{t}+
\frac{2P_{0}}{R_{0}}\eta
+\frac{ 3 P_{0}}{R_{0}^{2}}\eta^{2}=0,\\
\\
\varkappa=\frac{2R_{0}^{2}P_{0}}{3\chi Nu (n-1) T_{0}}\,.
    \end{gathered}
  \label{eq: state_eq}
\end{equation}

From Eq.(\ref{eq: density}) we also have the dependence $\rho$ on $\eta$ by means of formula (\ref{eq: rel1})
\begin{equation}
\begin{gathered}
\rho=\rho_{0}-\mu \eta +\mu_{1} \eta^{2}, \quad
\rho_{0}=\frac{\rho_{l}}{1-X+V_{0}\rho_{l}},\\
\\
\mu=\frac{3\rho_{l}^{2}V_{0}}{R_{0}(1-X+V_{0}\rho_{l})^{2}}, \quad
\mu_{1}=\frac{6 \rho_{l}^{2} V_{0}(2 \rho_{l}
V_{0}-1+X)}{R_{0}^{2}(1-X+\rho_{l} V_{0})^{3}}.
 \label{eq: desity1}
\end{gathered}
\end{equation}

We use the system of equations for description of the motion of a gas-liquid mixture flow  in the form
\begin{equation}
  \begin{gathered}
  \frac{\partial \rho}{\partial t} + \frac{\partial (\rho\,u)}{\partial x} = 0,  \quad
  \rho\left(\frac{\partial u}{\partial t} + u\,\frac{\partial u}{\partial x}\right)
  +  \frac{\partial P}{\partial x}=0,   \hfill
  \end{gathered}
  \label{eq: continuity_and_E_eq}
\end{equation}
where $u=u(x,t)$ is a velocity of a flow of a gas-liquid mixture.

Eq.(\ref{eq: continuity_and_E_eq}) together with Eqs.(\ref{eq:
state_eq}) and \eqref{eq: desity1} can be applied for describing  nonlinear waves in a gas-liquid medium. This system of equations can be written as
\begin{equation}
  \begin{gathered}
  \eta_{t} + u \eta_{x}+\eta u_{x} -\frac{\rho_{0}}{\mu} u_{x} -\frac{2 \mu_{1}}{\mu} \eta \eta_{t} = 0,  \hfill \\
  \\
  -\frac{\rho_{0}}{\mu}\left(u_{t}+u u_{x}\right)+\eta u_{t}-\frac{1}{\mu} P_{x}=0,   \hfill \\
  \\
  P+\frac{\eta}{R_{0}}P+\frac{3n\varkappa}{R_{0}}P\eta_{t}+\varkappa
  P_{t}= -\varkappa \rho_{l}(R_{0}+\eta)\eta_{ttt}-\hfill\\
  -(3n+4) \varkappa \rho_{l} \eta_{t}
  \eta_{tt}-\frac{\rho_{l}(3R_{0}^{2}+4\nu\varkappa)}{3R_{0}}\eta_{tt}-
  \frac{\rho_{l}(6R_{0}^{2}-4\nu\varkappa)}{3R_{0}^{2}}\eta
  \eta_{tt}-\hfill
  \\
  -\frac{\rho_{l}(8\nu\varkappa(3n-1)+9R_{0}^{2})}{6R_{0}^{2}}\eta_{t}^{2}
  -\frac{4\nu\rho_{l}}{3R_{0}}\eta_{t}-
  \frac{2P_{0}}{R_{0}}\eta+
   \frac{3P_{0}}{R_{0}^{2}}\eta^{2}+P_{0}. \hfill
  \end{gathered}
  \label{eq: main system}
\end{equation}

Consider the linear case of the system of equations (\ref{eq: main system}). Assuming, that pressure in a mixture is proportional to perturbation radius,  we obtain  the linear wave equation for the radius perturbations from Eqs.\eqref{eq: main system}
\begin{equation}
\eta_{tt}=c_{0}^{2} \, \eta_{xx}, \quad c_{0}=\sqrt{\frac{3P_{0} }{\mu R_{0}}}.
\end{equation}

Let us introduce the following dimensionless variables
\begin{equation*}
\begin{gathered}
  t = \frac{ l }{ c_{0} }\, t', \quad   x = l\, x', \quad  u = c_{0}\, u', \quad
  \eta=R_{0} \eta^{'}, \quad P = P_0\, P'+P_0,
  \label{eq: non-dim_subst}
\end{gathered}
\end{equation*}
where $l$ is the characteristic length of wave.

Using the dimensionless variables the system of equations (\ref{eq: main system}) can be reduced  to the following (the primes are omitted)
\begin{equation}
  \begin{gathered}
  \eta_{t}-\frac{\rho_{0}}{\mu R_{0}}u_{x} + u \eta_{x}+\eta u_{x}-\frac{2\mu_{1} R_{0}}{\mu} \eta \eta_{t}= 0,  \hfill \\
  \\
  -\frac{\rho_{0}}{\mu R_{0}}\left(u_{t}+u u_{x}\right)+\eta u_{t}- \frac{1}{3} P_{x}=0,   \hfill \\
  \\P+\varkappa_{1} P_{t}+\eta P +3n\varkappa_{1}\,\eta_{t} P=-\gamma \eta_{ttt}-\gamma \eta \eta_{ttt}-(3n+4)\gamma\eta_{t} \eta_{tt}-
  \\  -(\beta_{1}+\beta_{2})\,\eta_{tt}- (2\beta_{2}-\beta_{1})\, \eta\eta_{tt}-\left(\frac{3n-1}{2}\beta_{1}+\frac{3}{2}\beta_{2}\right)\eta_{t}^{2}
  -\\
  -\lambda \eta_{t}-3 \eta+3\eta^{2},  \hfill
  \end{gathered}
  \label{eq: main non-dim system}
\end{equation}
where the parameters are determined by formulae
\begin{equation}
  \begin{gathered}
    \lambda = \frac{ 4 \nu \rho_{l}c_{0} }{ 3 P_0\, l }+3n\varkappa_{1},\quad
     \beta_{1}=\frac{4\nu\,\varkappa\,\rho_{l}\, c_{0}^2}{3\,P_{0}\,l^{2}},\quad
    \beta_{2} = \frac{\rho_{l}\, c_{0}^2\,R_{0}^2}{P_{0}\,l^{2}},\\
    \gamma=\frac{\varkappa\rho_{l} R_{0}^{2}c_{0}^{3}}{P_{0}l^{3}},\,\quad
    \varkappa_{1}=\frac{\varkappa c_{0}}{l}.
  \end{gathered}
  \label{eq: non-dim_parameters}
\end{equation}

Numerical and analytical task by means of  the system of equations \eqref{eq: main non-dim system} is a difficult problem.

\section{Basic nonlinear evolution equation for describing nonlinear waves in liquid with bubbles}

We can not find the exact solutions of the system of nonlinear differential equations (\ref{eq: main non-dim system}). To account for the slow variation of the waveform, we introduce a scale transformation of independent variables  \cite{Kudr_2006a}
\begin{equation}
\begin{gathered}
  \xi = \varepsilon^m(x-t), \quad \tau= \varepsilon^{m+1}\, t ,\quad
  m > 0, \quad \varepsilon \ll 1,
  \label{eq: rescaling_coordinates}
    \end{gathered}
\end{equation}

\begin{equation*}
  \frac{\partial}{\partial x} = \varepsilon^m \frac{\partial}{\partial \xi},\quad
  \frac{\partial}{\partial t} = \varepsilon^{m+1} \frac{\partial}{\partial \tau}
  - \varepsilon^m \frac{\partial}{\partial \xi}.
\end{equation*}

Substituting (\ref{eq: rescaling_coordinates}) into (\ref{eq: main
non-dim system}) and dividing  on $\varepsilon^{m}$  in first two equations we have the following system of equations

\begin{equation}
  \begin{gathered}
    \varepsilon \eta_{\tau} - \eta_{\xi} -\frac{\rho_{0}}{\mu R_{0}} u_{\xi}
      +\, \eta u_{\xi}
      +\, u \eta_{\xi} -\varepsilon \frac{2\mu_{1} R_{0}}{\mu} \eta \eta_{\tau}
      + \, \frac{2\mu_{1} R_{0}}{\mu} \eta \eta_{\xi} = 0
\end{gathered}
  \label{eq: general_rescaled_system1}
\end{equation}

\begin{equation}
  \begin{gathered}
    \varepsilon \left(-\frac{\rho_{0}}{\mu R_{0}}\right) u_{\tau} +\frac{\rho_{0}}{\mu R_{0}} u_{\xi}
    +\varepsilon \eta u_{\tau} - \, \eta u_{\xi}
      - \, \frac{\rho_{0}}{\mu R_{0}} u u_{\xi}
      - \frac{1}{3}\,  P_{\xi} = 0
\end{gathered}
  \label{eq: general_rescaled_system2}
\end{equation}

\begin{equation}
  \begin{gathered}
    P +\varepsilon^{m+1} \varkappa_{1} P_{\tau}-\varepsilon^{m} \varkappa_{1} P_{\xi}
     +\, \eta P+\varepsilon^{m+1} 3n\varkappa_{1} \eta_{\tau} P-\varepsilon^{m} 3n\varkappa_{1} \eta_{\xi} P=\hfill \vspace{0.1cm} \\=
      -\varepsilon^{3m+3}  \gamma \eta_{\tau\tau\tau}+\varepsilon^{3m+2} 3\gamma \eta_{\tau\tau\xi}-
      \varepsilon^{3m+1} 3\gamma \eta_{\tau\xi\xi}\,+
      \varepsilon^{3m} \gamma \eta_{\xi\xi\xi}-\hfill \vspace{0.2cm} \\
      -\varepsilon^{3m+3}  \gamma \eta \eta_{\tau\tau\tau}+\varepsilon^{3m+2} 3\gamma \eta \eta_{\tau\tau\xi}-
      \varepsilon^{3m+1} 3\gamma \eta \eta_{\tau\xi\xi}+ \varepsilon^{3m} \gamma \eta \eta_{\xi\xi\xi}-\hfill \vspace{0.2cm} \\
      -\varepsilon^{3m+3} (3n+4) \gamma \eta_{\tau} \eta_{\tau\tau}
      + \varepsilon^{3m+2} (3n+4) \gamma \eta_{\xi} \eta_{\tau\tau}+\hfill \vspace{0.2cm} \\
      +\varepsilon^{3m+2} 2(3n+4) \gamma \eta_{\tau} \eta_{\tau\xi}
      -\varepsilon^{3m+1} 2(3n+4) \gamma \eta_{\xi} \eta_{\tau\xi}-\hfill \vspace{0.2cm} \\
      -\varepsilon^{3m+1} (3n+4) \gamma \eta_{\tau} \eta_{\xi\xi}
      +\varepsilon^{3m} (3n+4) \gamma \eta_{\xi} \eta_{\xi\xi}
      -\varepsilon^{2m+2} \,(\beta_{1}+\beta_{2})\, \eta_{\tau\tau}\,+\hfill \vspace{0.2cm} \\
      + \, \varepsilon^{2m+1} 2 \,(\beta_{1}+\beta_{2})\, \eta_{\tau\xi}\,
      - \, \varepsilon^{2m} \,(\beta_{1}+\beta_{2}) \, \eta_{\xi\xi}\,
      -\varepsilon^{2m+2} \,(2\beta_{2}-\beta_{1})\, \eta \eta_{\tau\tau}\,-\hfill \vspace{0.2cm} \\
      + \, \varepsilon^{2m+1} 2 \,(2\beta_{2}-\beta_{1})\, \eta \eta_{\tau\xi}\,
      - \, \varepsilon^{2m} \,(2\beta_{2}-\beta_{1}) \, \eta \eta_{\xi\xi}\,-\hfill \vspace{0.3cm} \\
      -\varepsilon^{2m+2}\left(\frac{3n-1}{2} \beta_{1}+\frac{3}{2}\beta_{2}\right)\eta_{\tau}^{2}
      +\varepsilon^{2m+1} 2\left(\frac{3n-1}{2} \beta_{1}+\frac{3}{2}\beta_{2}\right)\eta_{\tau}\eta_{\xi} -\hfill \vspace{0.3cm} \\-\varepsilon^{2m}\left(\frac{3n-1}{2} \beta_{1}+\frac{3}{2}\beta_{2}\right) \eta_{\xi}^{2}
      -\, \varepsilon^{m+1} \lambda \, \eta_{\tau}\,
      + \,   \varepsilon^{m}  \lambda\, \eta_{\xi}-3 \, \eta\, +\,3 \eta^{2}. \hfill
     \end{gathered}
  \label{eq: general_rescaled_system3}
\end{equation}

Now we assume that the variables $u$, $\eta$ and $P$ can be represented asymptotically as series in powers of $\varepsilon$ about an equilibrium state
\begin{equation}
  \begin{gathered}
    u = \varepsilon u_1 + \varepsilon^{2} u_2 +  \ldots,  \,\,\,
    \eta  =\varepsilon \eta _1 + \varepsilon^{2} \eta _2 +\ldots,  \,\,\,
    P = \varepsilon P_1  + \varepsilon^{2} P_2   +\ldots
  \end{gathered}
  \label{eq: asymptotic_expansion1}
\end{equation}

Substituting (\ref{eq: asymptotic_expansion1}) in (\ref{eq: general_rescaled_system1})-(\ref{eq: general_rescaled_system3}) and equating expressions at  $\varepsilon$, we have the equations
\begin{equation*}
  \begin{gathered}
    -\eta_{1\, \xi} -\frac{\rho_{0}}{\mu R_{0}} u_{1\, \xi} = 0\, ,\quad
    -\frac{\rho_{0}}{\mu R_{0}} u_{1\, \xi} + \frac{1}{3}\, P_{1\, \xi} = 0\, ,\quad
     P_{1} = -3 \, \eta_{1}.
  \end{gathered}
\end{equation*}

Solving this system of equation, we have
\begin{equation}
  \begin{gathered}
    \eta_{1}(\xi,\tau) = -\frac{\rho_{0}}{\mu R_{0}} u_{1}(\xi,\tau)\, +\, \psi(\tau)\, ,\quad
    P_{1}(\xi,\tau) = -3 \, \eta_{1}(\xi,\tau),
  \end{gathered}
  \label{eq: null_approximation_relations1}
\end{equation}
where $\psi(\tau)$  is an arbitrary function. We suppose  below that  $\psi (\tau )=0$.

Substituting (\ref{eq: asymptotic_expansion1}) in (\ref{eq: general_rescaled_system1})-(\ref{eq: general_rescaled_system3}) and equating factors at $\varepsilon^2$ to zero, we have the system of equations
\begin{equation}
\begin{gathered}\label{EE1}
  \eta_{1 \tau}-\eta_{2 \xi}- \frac{\rho_{0}}{\mu R_{0}} u_{2 \xi}+\eta _1 u_{1\xi }+  u_1
 \eta _{1\xi } +\frac{2\mu_{1} R_{0}}{\mu} \eta_{1} \eta_{1\xi}=0, \hfill \vspace{0.2cm}\\
  \\
 - \frac{\rho_{0}}{\mu R_{0}}  u_{1 \tau}+ \frac{\rho_{0}}{\mu R_{0}} u_{2 \xi} - \eta_{1} u_{1\xi} -\frac{\rho_{0}}{\mu R_{0}} u_1 u_{1\xi } - \frac{1}{3}\, P_{2\, \xi}=0, \hfill \vspace{0.2cm} \\
 \\
 P_{2} -\varepsilon^{m-1} \varkappa_{1} P_{\xi}+\eta_{1} P_{1}-\varepsilon^{m} 3n\varkappa_{1} \eta_{1\xi} P_{1}=  \varepsilon^{3m-1} \gamma \eta_{1\xi\xi\xi}+\varepsilon^{3m} \gamma \eta_{1} \eta_{1\xi\xi\xi} +\hfill \vspace{0.2cm} \\
+\varepsilon^{3m}(3n+4) \gamma \eta_{1\xi} \eta_{1\xi\xi} -\varepsilon^{2m-1} \,(\beta_{1}+\beta_{2}) \, \eta_{1\xi\xi}-\varepsilon^{2m} \,(2\beta_{2}-\beta_{1}) \, \eta_{1} \eta_{1\xi\xi}-\hfill \vspace{0.2cm} \\
-\varepsilon^{2m}\left(\frac{3n-1}{2} \beta_{1}+\frac{3}{2}\beta_{2}\right) \eta_{1\xi}^{2}+
 \varepsilon^{m-1}  \lambda\, \eta_{1\xi}-3\eta_{2}+3\eta_{1}^{2}. \hfill
 \end{gathered}
\end{equation}

Taking Eqs.\eqref{eq: null_approximation_relations1} and \eqref{EE1}
into account we have the equation for pressure $P_1$ in the form
\begin{equation}
\begin{gathered}
 P_{1\tau}\,+\alpha\,P_{1}P_{1 \xi}+
 \varepsilon^{2m-1}\, \frac{\beta_{1}+\beta_{2}}{6}\, P_{1\xi\xi\xi}\,
-\varepsilon^{2m}\, \frac{2\beta_{2}-\beta_{1}}{18}\, P_{1}
P_{1\xi\xi\xi}\,- \hfill \\- \varepsilon^{2m}\,
\left(\frac{3n-2}{18}\,\beta_{1}+\frac{5}{18}\,\beta_{2}\right)\,
P_{1\xi} P_{1\xi\xi}\, =
\varepsilon^{m-1} \, \left(\frac{\lambda}{6}\, - \, \frac{\varkappa_{1}}{2}\right) \,
P_{1\xi\xi} +\hfill \vspace{0.1cm} \\
+\varepsilon^{m} \, \frac{n\varkappa_{1}}{2}
\,\left(P_{1}P_{1\xi}\right)_{\xi}+ \varepsilon^{3m-1} \,
\frac{\gamma}{6}\, P_{1\xi\xi\xi\xi}-\varepsilon^{3m}
\,\frac{\gamma}{18} P_{1} P_{1\xi\xi\xi\xi}-\hfill \vspace{0.1cm} \\-
\varepsilon^{3m} \, \frac{(3n+5) \gamma}{18} P_{1\xi}
P_{1\xi\xi\xi}- \varepsilon^{3m} \, \frac{(3n+4) \gamma}{18} \,
P_{1\xi\xi}^{2}\,, \hfill \vspace{0.1cm} \\
\\
\alpha=\frac{3\mu R_{0}}{ \rho_{0}}-\frac{3\mu_{1}R_{0}}{\mu}+\frac{2}{3}.
 \label{eq:master equation1}
  \end{gathered}
\end{equation}

Assuming $m=1$ we obtain
the Burgers equation from Eq.\eqref{eq:master equation1}  when $\varepsilon \rightarrow 0$
\begin{equation}\begin{gathered}\label{EE2}
P_{1\tau}\,+\alpha\,P_{1}P_{1 \xi} =
\, \left(\frac{\lambda}{6}\, - \, \frac{\varkappa_{1}}{2}\right) \,
P_{1\xi\xi}.
\end{gathered}\end{equation}

Many solutions of the Burgers equation are well known \cite{Whitham, Kudr_2009aa, Kudr_2009bb}. By means of the Cole -- Hopf transformation \cite{Hopf50, Cole50} the Burgers equation can be transformed to the linear heat equation. The Cauchy problem for this equation can be solved \cite{Whitham}.

In the case $m=\frac12$, $\left(\frac{\lambda}{6}\, - \, \frac{\varkappa_{1}}{2}\right)\simeq \varepsilon^2$  we have
the famous Korteweg -- de Vries equation \cite{Korteweg} at  $\varepsilon \rightarrow 0$
\begin{equation}\begin{gathered}\label{EE3}
P_{1\tau}\,+\alpha\,P_{1}P_{1 \xi}+
\, \frac{\beta_{1}+\beta_{2}}{6}\, P_{1\xi\xi\xi}
=0.
\end{gathered}\end{equation}

This equation has the soliton solutions \cite{Zabuski}. The Cauchy problem for the Korteweg -- de Vries equation can be solved by the inverse scattering transform \cite{Gardner}. There are many methods to find exact solutions of the Korteweg -- de Vries equation \cite{Hirota, Kudryashov}. Unfortunately some of these approaches lead to the erroneous solutions \cite{Kudr_2009a, Kudr_2009b, Kudr_2009c}.

Assuming $m=\frac12$, $\left(\frac{\lambda}{6}\, - \, \frac{\varkappa_{1}}{2}\right)\simeq \varepsilon$  we have
the Korteweg -- de Vries -- Burgers equation at  $\varepsilon \rightarrow 0$
\begin{equation}\begin{gathered}\label{EE4}
P_{1\tau}\,+\alpha\,P_{1}P_{1 \xi}+
\, \frac{\beta_{1}+\beta_{2}}{6}\, P_{1\xi\xi\xi}\,
=
\, \left(\frac{\lambda}{6}\, - \, \frac{\varkappa_{1}}{2}\right)P_{1\xi\xi}
\end{gathered}\end{equation}

In the case $m=\frac13$, ${\beta_{1}+\beta_{2}}{}\simeq 6\, \varepsilon^{\frac13}$, $\left(\frac{\lambda}{6}\, - \, \frac{\varkappa_{1}}{2}\right)\simeq \,\varepsilon^{\frac23}$,  we have
the famous Kuramoto -- Sivashinsky equation \cite{Kuramoto, Sivashinsky83} when $\varepsilon \rightarrow 0$
\begin{equation}\begin{gathered}\label{EE5}
P_{1\tau}\,+\alpha\,P_{1}P_{1 \xi}+
\, P_{1\xi\xi\xi}=P_{1\xi\xi}+\frac{\gamma}{6}\,P_{1\xi\xi\xi\xi}.
\end{gathered}\end{equation}

The cauchy problem for equation \eqref{EE5} can not be solved by the inverse scattering transform but this equation has some exact solutions. The Kuramoto - Sivashinsky equation was studied in many papers (see, for the example \cite{Benney66,  Cohen76, Shkadov77, Topper78, Kudr_88, Michelson90, Kudr_90, Kudr_90a,  Kudr_91, Zhu96, Berloff97, Fu05, Wazwaz06, Kudr_07, Qin08, Kudr_08, Kudr_2009d,  Kudr_10}).

From Eq.\eqref{eq:master equation1} one can find some other nonlinear evolution equations for describing waves in a mixture of liquid and gas bubbles.

\section{Fourth order equation for describing wave processes in liquid with gas bubbles}

Assuming $m=\frac{1}{3}$, ${\lambda}=6\,\varepsilon^{2/3} \lambda^{'},$  $\lambda^{'} \sim 1$, ${\varkappa_{1}}=2\,\varepsilon^{2/3} \varkappa_{1}^{'}$, $\varkappa_{1}^{'} \sim 1$,
${\beta_{1}}={6}\varepsilon^{1/3}\,\beta_{1}^{'}$, $\beta_{1}^{'} \sim 1$, ${\beta_{2}}={6}\varepsilon^{1/3}\,\beta_{2}^{'}$,
$ \beta_{2}^{'} \sim 1$,\, $\delta>0$ in Eq.\eqref{eq:master equation1} and using transformations $P_{1}=\dfrac{1}{\varepsilon}\, P_{1}^{'}$,$\alpha=\varepsilon \, \alpha^{'}$
we have nonlinear evolution equation of the fourth order in the form
\begin{equation}
\begin{gathered}
P^{'}_{1\tau}+\alpha^{'} \, P^{'}_{1}P^{'}_{1\xi}+\left(\beta_{1}^{'}+\beta_{2}^{'}\right)\,P^{'}_{1\xi\xi\xi}-
\left(\frac{2\beta_{2}^{'}-\beta_{1}^{'}}{3}\right)\, P^{'}_{1}
P^{'}_{1\xi\xi\xi}\,-\hfill \\-\left(\frac{3n-2}{3}\beta_{1}^{'}+\frac{5}{3}\beta_{2}^{'}\right)
\, P^{'}_{1\xi} P^{'}_{1\xi\xi} =\left(\lambda^{'}-\varkappa'_{1}\right) P^{'}_{1
\xi\xi}+ n\varkappa'_{1} \,
\left(P^{'}_{1\xi} P^{'}_{1}\right)_{\xi}+\hfill \\+\frac{\gamma}{6}\, P_{1\xi\xi\xi\xi} -
\frac{\gamma}{18} P^{'}_{1} P^{'}_{1\xi\xi\xi\xi}-
\frac{(3n+5) \gamma}{18} \, P^{'}_{1\xi} P^{'}_{1\xi\xi\xi}
-  \frac{(3n+4) \gamma}{18} \, P^{'2}_{1\xi\xi}.\hfill
\label{eq: extKS1}
\end{gathered}
\end{equation}

Coefficient $\gamma/6$ at the fourth derivative in \eqref{eq: extKS1} can be presented in the form
\begin{equation}
\Gamma=\frac{\gamma}{6}=\frac{4\,\rho_{l}\,c_{0}^{3}\,R_{0}^{4}}{9\chi_{g}\,T_{0}\,Nu\,
(n-1)\,l^{3}}.
\label{c3}
\end{equation}

We can see that the value $\Gamma$ in \eqref{c3} depends on the initial radius of a bubble, sound velocity, coefficient of the heat transfer and the Nusselt mumber. The heat capacity of liquid is much more than the the heat capacity of the gas in bubble. As consequence the temperature of liquid is not changed but the gas in bubble can be heated and cooled. In the case of the isothermal processes we have $Nu \rightarrow \infty$ and Eq.\eqref{eq: extKS1} becomes the third order equation.

Let us look for the exact solutions of equation \eqref{eq: extKS1}.
Using the variables
\begin{equation*}
\xi=\frac{\gamma}{6(\beta_{1}^{'}+\beta_{2}^{'})}\xi^{'},\quad
\tau=\frac{\gamma^{3}}{216(\beta_{1}^{'}+\beta_{2}^{'})^{4}}\tau^{'},\quad
P^{'}_{1}=\frac{3(\beta_{1}^{'}+\beta_{2}^{'})}{2\beta_{2}^{'}-\beta_{1}^{'}} v.
\end{equation*}
 we obtain the equation (the primes are omitted)
\begin{equation}
\begin{gathered}
v_{\tau}+\alpha_{1} \, v\,v_{\xi}+v_{\xi\xi\xi}- ( v v_{\xi\xi})_{\xi}\, -
\beta \, v_{\xi} v_{\xi\xi} =\\
\hfill=\sigma \, v_{\xi\xi}+ \sigma_{1} \,
\left(v v_{\xi} \right)_{\xi}+ v_{\xi\xi\xi\xi} -
\gamma_{2} (v v_{\xi\xi\xi})_{\xi}
-\gamma_{3} \left(v_{\xi} v_{\xi\xi}\right)_{\xi}, \hfill
\label{eq: extKS1_tr}
\end{gathered}
\end{equation}
where parameters of Eq.\eqref{eq: extKS1_tr} can be written in the form
\begin{equation*}
\begin{gathered}
\alpha_{1}=\frac{\alpha'\gamma^{2}}{12(\beta_{1}^{'}+\beta_{2}^{'})^{2}(2\beta_{2}^{'}-
\beta_{1}^{'})},\quad
\beta=\frac{(3n-1)\beta_{1}^{'}+3\beta_{2}^{'}}{2\beta_{2}^{'}-\beta_{1}^{'}},\quad
\sigma=\frac{\gamma\left(\lambda^{'}-\varkappa'_{1}\right)}{6(\beta_{1}^{'}+\beta_{2}^{'})^
{2}},\vspace{2mm} \\
\sigma_{1}=\frac{n\varkappa'_{1}\,\gamma}{2(2\beta_{2}^{'}-\beta_{1}^{'})(\beta_{1}^{'}+
\beta_{2}^{'})},\quad
\gamma_{2}=\frac{\beta_{1}^{'}+\beta_{2}^{'}}{2\beta_{2}^{'}-\beta_{1}^{'}},\quad
\gamma_{3}=\frac{(3n+4)(\beta_{1}^{'}+\beta_{2}^{'})}{2\beta_{2}^{'}-\beta_{1}^{'}}.
\end{gathered}
\end{equation*}

Using traveling wave ansatz $v(\xi,\tau)=y(z), z=\xi-C_{0}\tau$ and integrating the equation on  $z$, we have
\begin{equation}
\begin{gathered}
C_{1}-C_{0}\, y +\frac{\alpha_{1}}{2}\, y^{2}+y_{zz} -y y_{zz} -\frac{\beta}{2}\, y_{z}^{2}-\sigma\, y_{z}-\sigma_{1}\, y y_{z}-\\
-y_{zzz}+\gamma_{2}\, y y_{zzz}+\gamma_{3}\, y_{z} y_{zz}=0.
\end{gathered}
\label{eq: reduced_extKS1}
\end{equation}
To find the solution of \eqref{eq: reduced_extKS1} we use the simplest equation method  \cite{Kudr_05,Kudr_05a, Kudr_08a}
\begin{equation}
y(z)=a_{0}+\,w,\quad w\equiv w(z),
\label{tr1}
\end{equation}
where $w(z)$ is the solution of the linear equation
\begin{equation}
w_{zz}=A\,w_{z}+B\,w.
\label{eq: s_eq}
\end{equation}

Substituting the transformation \eqref{tr1} and taking into account Eq.\eqref{eq: s_eq}  we obtain the following solution of the equation \eqref{eq: reduced_extKS1}
\begin{equation}
\begin{gathered}
y(z)=a_0 + C_{3} {\rm e}^{kz} +
C_{2}{\rm e}^{-kz},
\label{sl4}
\end{gathered}
\end{equation}
\begin{equation*}
\begin{gathered}
k= \frac { \left(\beta( \gamma_{3}+\gamma_{2} )+\sqrt {{\beta}^{2} (\gamma_{3}+\gamma_{2})(\gamma_{3}-3\,\gamma_{2})
+8\,\gamma_{3}(\gamma_{3}+\gamma_{2})(\beta+2\,\sigma_{1}\,\gamma_{3}) } \right) }{4\gamma_{3}\, \left( \gamma_{3}+\gamma_{2} \right) },
\end{gathered}
\end{equation*}
where $C_{2}$ and $C_{3}$ are arbitrary constants, parameters $a_0$, $A$, $B$, $\alpha_1$ $C_0$ and $C_1$ are
determined by formulae
\begin{equation*}
\begin{gathered}
a_{0}=\frac {2\,\beta\,(\gamma_{2}+\gamma_{3}-1)
-4\,\sigma\,\gamma_{3} (\gamma_{3}+\gamma_{2})
-4\,\sigma_{1}\,\gamma_{3}-\beta^{2}}{2\,\gamma_{3}\,\beta+4\,\sigma_{1}\,\gamma_{3}^{2}-
\gamma_{2}\,\beta^{2}},
\vspace{0.3cm} \\
A=\frac {\beta}{2 \gamma_{3}}, \quad
B=\frac {-\gamma_{2}\,{\beta}^{2}+2\,\gamma_{3}\,\beta+4\,\sigma_{1}
\,\gamma_{3}^{2}}{4 \left( \gamma_{3}+\gamma_{2} \right) \gamma_{3}^{2}},
\vspace{0.3cm}\\
\alpha_{1}=\frac {4\,\gamma_{3}^{2}\beta+\gamma_{2}^{2}{\beta}^{3}-4\,\gamma_{2}\,{\beta}^{2}\gamma_{3}+8\,\sigma_{1}\,\gamma_{3}^{3}-
4\,\sigma_{1}\,\gamma_{3}^{2}\beta\,\gamma_{2}}{4\gamma_{3}^{3}
\left( \gamma_{3}+\gamma_{2} \right) },
\vspace{0.3cm}\\
C_{0}=\frac{1}{4\gamma_{3}^{3} \left( \gamma_{3}+\gamma_{2} \right) }\, \left(4\,\sigma_{1}\,\gamma_{3}^{3}-4\,\sigma_{1}\,\gamma_{3}^{2}-2\,
\sigma_{1}\,\gamma_{3}^{2}\beta+2\,\sigma_{1}\,\beta\,\gamma_{2}\,\gamma_{3}
+\gamma_{2}\,{\beta}^{2}+\right. \\ \left.+2\,\sigma\,\gamma_{3}^{2}\beta\,\gamma_{2}--4\,\gamma_{3}^{3}\sigma-4\,\gamma_{3}^{2}\gamma_{2}\,
\sigma+4\,\gamma_{3}^{2}\beta+2\,\beta\,\gamma_{2}\,\gamma_{3}-2
\,{\beta}^{2}\gamma_{3}-2\,\gamma_{2}\,{\beta}^{2}\gamma_{3}+\right. \\ \left. +2\,
\gamma_{3}\,\sigma\,\beta\,\gamma_{2}^{2}-2\,\gamma_{3}\,\beta+{
\beta}^{3}\gamma_{2}-\gamma_{2}^{2}{\beta}^{2}\right)\,,
\vspace{0.3cm}\\
C_{1}=\frac { \left( \beta-2\,\gamma_{3} \right)  \left( \beta^{2}-2\,\beta\,\gamma_{2}+2\,\beta-2\,\gamma_{3}\,\beta+
4\,\gamma_{2}\,\sigma\,\gamma_{3}+4\,\sigma_{1}\,\gamma_{3}+4\,\sigma\,\gamma_{3}^{2} \right) }{8\gamma_{3}^{3} \left( \gamma_{3}+\gamma_{2} \right) }.
\label{coeff_3}
\end{gathered}
\end{equation*}

For finding the periodic solutions of Eq.\eqref{eq: reduced_extKS1} we  use the expression in the form
\begin{equation}
y(z)=a_{0}+a_{1}\,w+a_{2}\,w^{2}+
a_{3}\,w^{3},\quad w\equiv w(z)
\label{transform1}
\end{equation}
where $w(z)$ also satisfies linear ordinary differential equation \eqref{eq: s_eq}.

Substituting Eq.\eqref{transform1} into Eq.\eqref{eq: reduced_extKS1} and equating the different expressions at $w(z)$ to zero, we have the system of the algebraic equations> Solving this system of equations we obtain the following values of parameters
\begin{equation}
\begin{gathered}
a_{2}=a_{1}=0,\quad a_{0}=1, \quad \gamma_{2}=1, \quad \gamma_{3}=-\frac{1}{3},
\quad B=-\frac{\alpha_{1}}{6(A-1)} \\
\sigma=-\frac {-2\,A^{2}+A-\alpha_{1}+A^{3}}{A-1}, \quad \sigma_{1}=\frac {-2\,A^{2}+A-\alpha_{1}+A^{3}}{A-1} \\
\beta=\frac{10\,A-4}{3},\quad C=0 ,\quad C_{0}=\alpha_{1}, \quad C_{1}=\frac{\alpha_{1}}{2}
\label{coeff_1}
\end{gathered}
\end{equation}
and solution of equation \eqref{eq: reduced_extKS1} in the form
\begin{equation}
\begin{gathered}
y(z)=1+ a_{3}\, \left[ C_{3}
\exp\left(z\,\frac{A}{2}+\frac{z\,\sqrt{p}}{6(A-1)}\right)
+C_{2}
\exp\left(z\,\frac{A}{2}-\frac{z\,\sqrt{p}}{6(A-1)}\right)
\right] ^{3}\\
 p=3 \left( A-1 \right)  \left( 3\,{A}^{3}-3\,{A}^{2}-2\,\alpha_{1} \right),\quad A\neq0,\quad A\neq1,\quad a_{3}\neq0
\label{sl1}
\end{gathered}
\end{equation}
In the case of
\begin{equation*}
\begin{gathered}
\frac { \left( -3\,A+3\,{A}^{2}+\sqrt {p} \right)}{6(A-1)}<0,\quad \frac { \left( -3\,A+3\,
{A}^{2}-\sqrt {p} \right)}{6(A-1)}<0
\end{gathered}
\end{equation*}
solution  \eqref{sl1} is finite function.

Consider case  $A=0$. In this case the parameters of equation and coefficients of expansion take the form
\begin{equation*}
\begin{gathered}
a_{1}=a_{2}=0,\quad a_{0}=1, \quad \gamma_{2}=1, \quad \gamma_{3}=-\frac{1}{3},
\quad B=\frac{\alpha_{1}}{6} \\
\sigma=-\alpha_{1}, \quad \sigma_{1}=\alpha_{1}, \quad
\beta=-\frac{4}{3},\quad C=0 ,\quad C_{0}=\alpha_{1}, \quad C_{1}=\frac{\alpha_{1}}{2}.
\label{coeff_2}
\end{gathered}
\end{equation*}
At $\alpha_{1}>0$ we have the following solution of equation \eqref{eq: reduced_extKS1}
\begin{equation}
\begin{gathered}
y(z)=1+ a_{3}\, \left( C_{3}\,{\rm e}^{\frac{\sqrt{6\alpha_{1}}z}{6} }+C_{2}\,{\rm e}^{-\frac{\sqrt{6\alpha_{1}}z}{6}}
 \right) ^{3},
a_{3}\neq0
\label{sl2}
\end{gathered}
\end{equation}
In the case of $\alpha_{1}<0$ the solution of equation \eqref{eq: reduced_extKS1} can be written as
\begin{equation}
\begin{gathered}
y(z)=1+ a_{3}\, \left( C_{3}\,\sin \left\{ \frac{\sqrt{6|\alpha_{1}|}z}{6} \right\} +C_{2}\,\cos \left\{ \frac{\sqrt{6|\alpha_{1}|}z}{6} \right\}  \right) ^{3}, \quad
 a_{3}\neq0.
\label{sl3}
\end{gathered}
\end{equation}

Solutions \eqref{sl2}, \eqref{sl3} of Eq.\eqref{eq: reduced_extKS1} have two arbitrary constants  $C_{2}$  and  $C_{3}$.

\begin{figure}[h]
\begin{center}
 \includegraphics[width=0.8\textwidth]{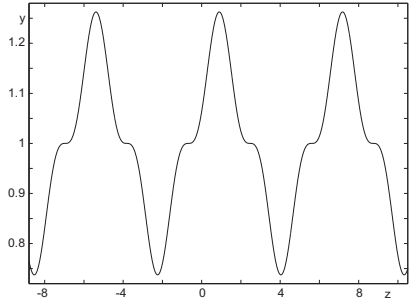}
   \caption{Periodic solution \eqref{sl3} of Eq.\eqref{eq: reduced_extKS1}}
\end{center}
 \label{fig1}
\end{figure}

Solutions \eqref{sl1}, \eqref{sl3} have physical sense. The first of them is limited at the values of parameters at conditions of the above mentioned. The solution \eqref{sl3} is the periodical solution. The velocity of the wave described \eqref{sl1}, \eqref{sl2}, \eqref{sl3}, is equal to parameter characterizing nonlinear transfer.

Dependence of a solution \eqref{sl3} of Eq.\eqref{eq: reduced_extKS1} with respect to  $z$ at values of parameters $\alpha_{1}=-1,\,C_{2}=0.04,\,C_{3}=0.05$  is given on Fig. 1.

\section{Conclusion}

We have studied the propagation of nonlinear waves in a mixture of liquid and gas bubbles taking into consideration the influence of heat transfer and viscosity.  We have applied different scales of time and coordinate and the method of perturbation to obtain nonlinear evolution equations for describing pressure waves. As a result we have found the nonlinear evolution equation
\eqref{eq:master equation1} of the fourth order for describing the pressure waves in a liquid with bubbles. This evolution equation allow us to take into account the influence of the viscosity and the heat transfer on the pressure waves. It is likely this nonlinear differential equation is new and generalizes many well -- known nonlinear equations as the Burgers equation, the Korteweg -- de  Vries equation, the Korteweg -- de Vries -- Burgers equation and the Kuramoto -- Sivashinsky equation.  Generally this equation is not integrable but this equation has some exact solutions that have been found in this paper.

\end{document}